\documentclass[aps,prl,reprint,superscriptaddress,nofootinbib]{revtex4-2}

\usepackage{amsmath,amssymb}
\usepackage{microtype}
\usepackage[hidelinks]{hyperref}

\newcommand{\I}{\mathrm I}
\newcommand{\ket}[1]{\lvert #1\rangle}
\newcommand{\bra}[1]{\langle #1\rvert}
\newcommand{\proj}[1]{\ket{#1}\!\bra{#1}}

\begin{document}

\title{When Complementary Measurements Count the Same Classical Bit Twice:\texorpdfstring{\\}{ }
Counterexamples to CQC, ECQC, and Complementarity-Based Certification}

\author{Jinbo Wang}
\affiliation{School of Mathematical Sciences, Peking University, Beijing 100871, China}
\author{Qihang Wang}
\affiliation{School of Mathematical Sciences, Peking University, Beijing 100871, China}
\author{Kun Chen}
\email{Contact author: chenkun@itp.ac.cn}
\affiliation{Institute of Theoretical Physics, Chinese Academy of Sciences, Beijing 100190, China}
\date{August 5, 2026}

\hypersetup{
  pdftitle={When Complementary Measurements Count the Same Classical Bit Twice: Counterexamples to CQC, ECQC, and Complementarity-Based Certification},
  pdfauthor={Jinbo Wang, Qihang Wang, and Kun Chen}
}

\begin{abstract}
Mutually unbiased measurements are commonly expected to expose independent
facets of a quantum state: a correlation that is classical in one basis should
disappear in a complementary basis.  In higher dimensions, however, this
intuition becomes particularly subtle because correlations recovered in
different settings need not represent different information.  To expose this
loophole, we propose a two-branch classical null test: before the setting is
chosen, a shared bit selects one of two orthogonal product preparations,
producing a rank-two classical--classical state, and the candidate protocol
then runs unchanged.  Different settings can read the same bit through
different outcome patterns.  This two-branch classical architecture disproves the complementary-quantum
correlation (CQC) conjecture in every dimension $d\geq3$.  A distinct
rank-two classical--classical state disproves its complete-basis extension
(ECQC) at $d=7$, with an overrun that grows without bound along prime
dimensions.  Its qutrit CQC instance also gives classical false positives for
a proposed quantum-correlation measure and a proposed one-sided
semi-device-independent steering criterion, and refutes a
conditional-probability conjecture.  The failures identify the missing
requirement: information read in different settings must be nonredundant.  In
experiments and applications, the same low-overhead architecture can serve as a
calibration test before a multibasis score is assigned quantum meaning.
\end{abstract}

\maketitle

\section{Introduction}
\label{sec:introduction}

Measuring a quantum system in one way can erase what was predictable in
another.  The extreme case is a pair of mutually unbiased bases (MUBs): a state
with a certain outcome in one basis gives every outcome with equal probability
in the other.  This familiar expression of complementarity suggests a simple
intuition: a classical source may coordinate Alice and Bob in one measurement
basis, but should lose that coordination when both switch to a complementary
basis.  Correlations observed in both bases are then taken as evidence of a
genuinely quantum shared resource.  The
complementary-quantum correlation (CQC) conjecture turns exactly this intuition
into an additive information budget: correlations accessible in two MUBs are
proposed to share a single budget set by the state's quantum mutual information
(QMI) \cite{Schneeloch2014}.  The same intuition appears in several concrete
proposals: the extended CQC (ECQC) conjecture for complete MUB sets
\cite{Iqbal2026}; a criterion that sums matched conditional probabilities in
complementary bases to detect entanglement \cite{Maccone2015}; and a claim that
residual complementary-basis correlations certify one-sided
semi-device-independent (1SSDI) steering \cite{Jebarathinam2025}.  It also
motivates complementary-basis measures of quantum correlation and MUB
entanglement witnesses \cite{Wu2014,Spengler2012}.  Experimentally, such
multibasis scores are attractive because they can be estimated from a few
local measurement settings, offering low-overhead routes to quantify
correlations and certify entanglement or steering without full state
tomography.

This intuition is natural for qubits but becomes subtle in higher dimensions.
In the simplest qubit encoding, a source maps the bit $K$ to the two $X$-basis
states.  The $X$ measurement reads $K$ exactly, while the mutually unbiased
$Z$ measurement gives the same $50{:}50$ distribution for both states and
erases it.  For $d\geq3$, however, there are many superpositions of
$\ket{x_1},\ldots,\ket{x_{d-1}}$.  Encode $K=0$ in $\ket{x_0}$ and $K=1$ in
any such superposition.  The binary test $X=0$ versus $X\neq0$ reads $K$
perfectly, while different choices for the $K=1$ state can have different
$Z$-basis statistics.  Choosing one whose $Z$ distribution differs from that
of $\ket{x_0}$ lets $Z$ retain partial information about the same $K$.  Any
theoretical criterion or experimental protocol that combines correlations
from complementary bases as evidence of a quantum resource must therefore
confront a common potential loophole: the same classical information may be
counted more than once.

To test this potential loophole, we propose a two-branch classical null test.
The test feeds a known classical control state $\rho_d$ into a protocol while
leaving its measurements and scoring rule unchanged.  If $\rho_d$ crosses a
threshold claimed to certify a quantum resource, the criterion has a classical
false positive.  For every $d\geq3$, construct $\rho_d$ as follows.  Let
$Z=\{\ket{j}\}$ be the computational basis and let $X$ be its Fourier MUB:
\begin{equation}
 \begin{aligned}
 \ket{u}&=\frac{1}{\sqrt d}\sum_{j=0}^{d-1}\ket{j},\qquad
 \ket{v}=\frac{\ket{0}-\ket{1}}{\sqrt2},\\
 \rho_d&=\frac12\proj{u}_A\otimes\proj{v}_B
       +\frac12\proj{v}_A\otimes\proj{u}_B .
 \end{aligned}
 \label{eq:state}
\end{equation}
To prepare $\rho_d$, sample a fair bit $K$ before choosing the measurement
setting.  For $K=0$, send $\ket u$ to Alice and $\ket v$ to Bob; for $K=1$,
swap them.  Thus $\rho_d$ is an equal mixture of two product preparations.
Because $\ket u$ and $\ket v$ are orthogonal, it is classical-classical and its
only shared correlation is $K$.  Moreover, $\ket u=\ket{x_0}$, so $X$
distinguishes the two preparations as $x=0$ versus $x\neq0$.  Their $Z$-basis
distributions also differ, allowing both bases to carry information about the
same $K$.

Applied without modification, this null source disproves CQC for every
$d\geq3$.  A distinct rank-two classical--classical state, specified in the
ECQC section, disproves ECQC at $d=7$ with an aggregate overrun that grows
without bound along prime dimensions \cite{Schneeloch2014,Iqbal2026}.  At
$d=3$, the same source violates the prediction that a complementary-basis
quantum-correlation measure vanishes on classical--classical (CC) states
\cite{Wu2014}, the claimed implication from that measure to 1SSDI steering
\cite{Jebarathinam2025}, and a proposed separable conditional-probability
interval \cite{Maccone2015}; a related full-rank $d=4$ source also defeats the
natural permutation-optimized repair of the latter.  Beyond these specific
failures, the same low-cost source provides a laboratory calibration: a
threshold crossing in an otherwise unchanged data pipeline directly diagnoses
classical overcounting.  Passing this null test is necessary but not
sufficient; a final certificate must still bound every classical or separable
source and every processing step allowed by its operational assumptions.

\section{Two-Branch Classical Null Test}
\label{sec:null}

For a bipartite state $\rho_{AB}$, the quantum mutual information (QMI) is
$\I(A:B)=S(\rho_A)+S(\rho_B)-S(\rho_{AB})$, where $S$ is the von Neumann
entropy.  By contrast, $\I(M^A:M^B)$ denotes the ordinary Shannon mutual
information between the classical outcomes when Alice and Bob measure a basis
$M$.  The states $\ket u$ and $\ket v$ in Eq.~\eqref{eq:state} are orthogonal,
so $\rho_d$ is diagonal in a local product basis.  It is therefore a rank-two
CC state: its correlation is a pre-existing classical label, and the state is
separable with zero discord in both directions.  The global state and both
marginals have nonzero spectrum $(1/2,1/2)$, giving $\I(A:B)=1$ bit.

The loophole appears when correlations from different settings are added.
For each setting $x$, data processing gives
$\I(A_x:B_x)\leq\I(A:B)=1$, but it does not bound the sum over $x$ by one bit.
Each setting is evaluated on a fresh copy of $\rho_d$, and both parties may
change their readout with $x$.  For our source, different MUB settings recover
the same $K$, so their sum counts one classical bit more than once.

No quantum information is cloned, and MUB geometry remains intact.  Mutual
unbiasedness fixes the exact outcome distribution obtained by measuring a
basis state in the other basis; it does not prevent different settings from
grouping their outcomes so as to reveal the same classical variable.
Multibasis correlations in nonproduct and zero-discord states were known
\cite{GuoWu2014,Kanjilal2018}; $\rho_d$ shows that this redundancy can cross
proposed quantum thresholds.  Nor does the construction contradict
information-exclusion bounds proved with one party's measurement fixed
\cite{Wu2009}: CQC and ECQC allow both parties to change their readout with the
setting.

\section{Counterexample to CQC}
\label{sec:cqc}

Let $Z=\{\ket{j}\}$ be the computational basis and
$X=\{\ket{x_k}\}$ its Fourier MUB,
$\ket{x_k}=d^{-1/2}\sum_j e^{2\pi i jk/d}\ket{j}$.  The
CQC conjecture asserts \cite{Schneeloch2014}
\begin{equation}
 \I(Z_A:Z_B)+\I(X_A:X_B)\leq\I(A:B).
 \label{eq:cqc}
\end{equation}
It was proved for pure states, states with one maximally mixed subsystem, and
all states when one measurement is minimally disturbing; extensive Monte
Carlo searches found no violation \cite{Schneeloch2014}.  Later work derived
a sufficient condition covering additional states and proposed an extension
to complete MUB families in prime dimensions \cite{Iqbal2026}.

Apply the null source in Eq.~\eqref{eq:state}.  In the $X$ basis,
$\ket{u}=\ket{x_0}$ always gives outcome $0$, whereas
$\ket{v}$ never does.  Thus the binary question ``zero or nonzero?''
identifies $K$ perfectly at either party, and
$\I(X_A:X_B)=1$ bit.  In the $Z$ basis the outcome laws of $\ket{u}$
and $\ket{v}$ are
\begin{equation}
 p=(1/d,\ldots,1/d),\qquad q=(1/2,1/2,0,\ldots,0),
\end{equation}
so that
\begin{equation}
 P_Z(j,k)=\tfrac12[p_jq_k+q_jp_k].
 \label{eq:pz}
\end{equation}
This distribution is nonproduct for every $d\geq3$, and its mutual
information is
\begin{equation}
 g_d=\frac{2}{d}\log_2\frac{8d}{(d+2)^2}
 +\frac{d-2}{d}\log_2\frac{2d}{d+2}>0.
 \label{eq:gd}
\end{equation}
The quantity $g_d$ is the residual $Z$-basis correlation generated by the same
bit $K$ already read perfectly in $X$; it is not a second shared variable.
Consequently,
$\I(X_A:X_B)+\I(Z_A:Z_B)=1+g_d>1=\I(A:B)$, disproving CQC for every
$d\geq3$.  For qutrits the measured sum is
\begin{equation}
 1+\frac13\log_2\frac{3456}{3125}
 =1.0484156759\ldots\ \text{bits}.
 \label{eq:qutrit}
\end{equation}
The violation is strict and therefore survives sufficiently weak white noise,
giving full-rank separable counterexamples.

\section{Counterexample to ECQC}
\label{sec:ecqc}

The complete-basis extension asks whether using all available complementarity
restores an additive information budget.  For a complete set $\mathcal M$ of
$d+1$ MUBs in prime dimension, ECQC sums the $d$ least-correlated bases---or,
equivalently, discards one largest term---and asserts \cite{Iqbal2026}
\begin{equation}
 \min_{\substack{\mathcal S\subset\mathcal M\\|\mathcal S|=d}}
 \sum_{M\in\mathcal S}\I(M^A:M^B)\leq\I(A:B).
 \label{eq:ecqc}
\end{equation}
For an odd prime $d$, let $\mathbb F_d$ denote integers modulo $d$.  Take
$M_\infty=Z$ and, for each $a\in\mathbb F_d$, define the quadratic basis
$M_a=\{\ket{a,b}:b\in\mathbb F_d\}$ by
\begin{equation}
 \ket{a,b}=\frac1{\sqrt d}\sum_{j=0}^{d-1}
 e^{2\pi i(aj^2+bj)/d}\ket{j},\qquad a,b\in\mathbb F_d.
 \label{eq:mubs}
\end{equation}
These $d+1$ bases form a complete MUB set.  For ECQC, use instead
\begin{equation}
 \ket{u_d^{\mathrm E}}=\frac{\ket0+\ket1}{\sqrt2},\qquad
 \ket{v_d^{\mathrm E}}=\frac{\ket0-\ket1}{\sqrt2},
 \label{eq:ecqc-branches}
\end{equation}
and let $\rho_d^{\mathrm E}$ be the equal swapped product mixture obtained
from Eq.~\eqref{eq:state} by replacing $\ket u,\ket v$ with
$\ket{u_d^{\mathrm E}},\ket{v_d^{\mathrm E}}$.  This is again a rank-two CC state with
$\I(A:B)=1$.  The computational basis gives the same outcome law for
$\ket{u_d^{\mathrm E}}$ and $\ket{v_d^{\mathrm E}}$, hence
$\I_{M_\infty}=0$.  In every quadratic MUB their laws are, up to a common
cyclic relabeling,
\begin{equation}
 p_j=\frac{1+c_j}{d},\qquad q_j=\frac{1-c_j}{d},\qquad
 c_j=\cos\frac{2\pi j}{d}.
 \label{eq:ecqc-laws}
\end{equation}
Their swapped branches give
\begin{equation}
 P^{(d)}_{jk}=\frac{1-c_jc_k}{d^2},\qquad
 J_d:=\I(P^{(d)}).
 \label{eq:ecqc-joint}
\end{equation}
Thus the $d+1$ scores are $\{0,J_d,\ldots,J_d\}$.  ECQC retains the zero term
and $d-1$ copies of $J_d$, giving $E_d=(d-1)J_d$.  At $d=7$,
\begin{equation}
 \begin{aligned}
 J_7&=0.2079841266\ldots,\\[-2pt]
 E_7&=1.2479047596\ldots>\I(A:B)=1.
 \end{aligned}
 \label{eq:d7}
\end{equation}
Appendices A--E supply the full derivations, exact
tables, robustness constructions, and independent executable checks.  In
particular, $J_d$ approaches a strictly positive
integral along odd primes.  Hence $d-1$ retained quadratic bases each
contribute an incomplete reading of the same branch bit, and $E_d$ grows
linearly along prime dimensions although the state contains only one bit of
QMI.  Complete MUB symmetry therefore amplifies redundant readouts into an
unbounded aggregate overrun.

\section{Certification Failures and Scope}
\label{sec:consequences}

\textit{Quantum-correlation measure.---}
The same qutrit state tests more than a QMI conjecture.  The Holevo correlation
upper-bounds how much information Bob can obtain about Alice's measurement
result from his conditional quantum states.  The measure $Q_2^{\rightarrow}$
maximizes this correlation over Alice's basis and then over a MUB, interpreting
the residual as ``genuine quantum'' correlation \cite{Wu2014}.  The proposal
predicts $Q_2^{\rightarrow}=0$ for classical--quantum (CQ) states, which include
CC states.  In Eq.~\eqref{eq:state}, measuring Alice in $X$ prepares orthogonal
branch states for Bob and achieves the one-bit optimum.  Yet the MUB $Z$ gives
\begin{equation}
 Q_2^{\rightarrow}\geq0.1908745046\ldots>0
 \label{eq:q2}
\end{equation}
on a CC state, directly violating that null property.

\textit{1SSDI steering.---}
Here steering asks whether the observed data can be explained by a bounded
hidden variable on one side and pre-existing local quantum states on the other.
Such an explanation is a local
hidden-variable/local-hidden-state (LHV--LHS) model.  A later work
claimed universally that $Q_2^{\rightarrow}>0$ demonstrates two-setting 1SSDI
steering \cite{Jebarathinam2025}.  Here the two product branches in
Eq.~\eqref{eq:state} are already an explicit two-valued LHV--LHS model, and its
hidden-variable dimension $2$ obeys the stipulated $2\leq d_A=3$.  The state
is therefore unsteerable in the stated model despite Eq.~\eqref{eq:q2}.

\textit{Conditional-probability entanglement score.---}
For a basis $M$, let $a_i$ and $b_i$ denote Alice's and Bob's outcomes assigned
the same label $i$.  The score $\mathcal S_M=\sum_i p(a_i|b_i)$ sums the
probabilities that Alice obtains the paired outcome $a_i$ conditioned on Bob
obtaining $b_i$.  It was conjectured that separable states obey
$1\leq\mathcal S_X+\mathcal S_Z\leq d+1$ \cite{Maccone2015}.  With the fixed
same-index pairing used in that definition, the qutrit tables of
Eq.~\eqref{eq:state} give $\mathcal S_X+\mathcal S_Z=0.8$.  The original work
noted that the pairing is arbitrary but did not optimize it.  Even the natural
repair that maximizes over a separate outcome permutation in each setting
fails at the upper end: an explicit full-rank separable $d=4$ state gives
$5.0710794\ldots>5$.  The constructions and exhaustive assignment check are
given in Appendix D.

\textit{Scope.---}
Four precise statements fail: the universal CQC and ECQC bounds, the claimed
CQ/CC-null property of $Q_2^{\rightarrow}$, its universal bridge to 1SSDI
steering, and the conjectured conditional-probability interval.  Where such a
statement was a theorem or certification implication, the counterexample shows
that its conclusion cannot hold universally.  Entropic-uncertainty relations,
the independently proved mutual-information entanglement criterion of
Ref.~\cite{Maccone2015}, and other MUB witnesses with independent separable
bounds \cite{Spengler2012,Coles2017} remain intact.  Nor is this a generic
attack on QKD, device-independent security, or finite-key analysis; a
cryptographic consequence requires its specified source, side-information,
and adversary assumptions.  Practically, the two-branch CC source can calibrate
an unchanged protocol before quantum, steering, or cryptographic meaning is
assigned to its score.  Failure exposes a classical false positive; passing is
only a necessary check, not a complete resource proof.

\section{Discussion}
\label{sec:discussion}

Complementarity constrains fine-grained questions asked of one quantum system,
but it does not stop separate experiments from rereading a classical common
cause.  This distinction is invisible in basis overlaps yet decisive for
additive scores.  Our null test makes it operational: multisetting
certification should first survive the same classical source across its entire
pipeline.  A complete resource claim must then bound all allowed classical or
separable sources; MUB geometry alone does not supply a common information
budget.

The dimensional boundary is informative.  For a qubit, the subspace
orthogonal to $\ket{x_0}$ is spanned by $\ket{x_1}$, which has the same uniform
$Z$ statistics as $\ket{x_0}$.  Our source therefore saturates rather than
violates CQC at $d=2$.  Earlier numerical studies found no two-qubit
counterexample \cite{Schneeloch2014}, including subsequent tests based on
sampling density matrices at fixed purity \cite{Alsing2022}.  These results
provided numerical evidence, but did not establish universal validity.

In subsequent companion work, we proved the CQC inequality for every
two-qubit state and every pair of mutually unbiased local bases on each
subsystem, without restrictions on rank, marginals, or separability
\cite{WangQubits2026}.  That proof combines a qubit information-exclusion
tradeoff with data processing under local dephasing.  Together with the
explicit counterexamples constructed here for every $d\geq3$, it establishes
a sharp dimensional boundary for the universal CQC inequality on
$\mathbb C^d\otimes\mathbb C^d$: it holds for all states at $d=2$, whereas
counterexamples exist in every dimension $d\geq3$.  The two works therefore
resolve the two-basis CQC question across equal local dimensions; the ECQC
and certification counterexamples remain separate conclusions of the present
work.

\begin{acknowledgments}
\textit{Acknowledgments.—} AI-assisted tools (OpenAI Codex, GPT-5.6-Sol) were used for exploratory search,
literature organization, and error checking.  The authors directed these uses,
independently verified all constructions, calculations, and source claims, and
take full responsibility for the scientific content.  Kun Chen acknowledges
support from the Strategic Priority Research Program of the Chinese Academy of
Sciences under Grant No.~XDB1680102.
\end{acknowledgments}

\textit{Data and code availability.—}
No experimental data were generated.  Independent executable certificates for
all numerical values and counterexamples are provided in the \texttt{anc/}
directory of the arXiv source archive.

\clearpage
\onecolumngrid
\appendix
\section*{Appendix A: The CC state and the CQC counterexample}

These appendices prove the CQC and ECQC counterexamples, derive the asymptotic
ECQC overrun, and give the explicit classical models and
conditional-probability constructions used in the main text.  All logarithms
are base two.  We write
$\I(A:B)=S(\rho_A)+S(\rho_B)-S(\rho_{AB})$ for the QMI of a state and
$\I(M^A:M^B)$ for the Shannon mutual information between the classical
outcomes of a local basis measurement $M$.

For $d\geq3$, let
\begin{equation}
 \ket{u}=\frac1{\sqrt d}\sum_{j=0}^{d-1}\ket{j},\qquad
 \ket{v}=\frac{\ket0-\ket1}{\sqrt2},
\end{equation}
and
\begin{equation}
 \rho_d=\frac12\proj{u}_A\otimes\proj{v}_B
       +\frac12\proj{v}_A\otimes\proj{u}_B.
 \label{eqs:rho}
\end{equation}
The local states are orthogonal because
$\langle u|v\rangle=(1-1)/\sqrt{2d}=0$.  Extending
$\{\ket u,\ket v\}$ to a local orthonormal basis makes
Eq.~\eqref{eqs:rho} diagonal in a product basis.  It is therefore
classical--classical (CC): its only correlation is the classical branch label,
and it is separable and zero-discord in both directions.  The two global
product states are orthogonal, and
\begin{equation}
 \rho_A=\rho_B=\tfrac12(\proj u+\proj v).
\end{equation}
Thus the nonzero spectra of $\rho_{AB}$, $\rho_A$, and $\rho_B$ are all
$(1/2,1/2)$, proving $\I(A:B)=1$ bit.

Let $Z$ be the computational basis and let
\begin{equation}
 \ket{x_k}=\frac1{\sqrt d}\sum_{j=0}^{d-1}\omega^{jk}\ket j,
 \qquad \omega=e^{2\pi i/d},
\end{equation}
define the Fourier basis $X$.  In $X$, the outcome laws of $\ket u$ and
$\ket v$ are, respectively,
\begin{equation}
 \delta_{k0},\qquad
 s_k=|\langle x_k|v\rangle|^2
 =\frac2d\sin^2\frac{\pi k}{d},\qquad s_0=0.
\end{equation}
The joint law is therefore
\begin{equation}
 P_X(j,k)=\tfrac12(\delta_{j0}s_k+s_j\delta_{k0}).
 \label{eqs:px}
\end{equation}
The binary coarse graining $j\mapsto\boldsymbol 1[j=0]$ recovers the branch
bit perfectly on both sides.  It gives one bit of classical mutual information.
Since a local measurement cannot exceed the QMI, this also proves directly that
$\I(X_A:X_B)=1$.

In $Z$, the two local outcome laws are
\begin{equation}
 p_j=\frac1d,\qquad q_j=\frac12(\delta_{j0}+\delta_{j1}),
\end{equation}
and
\begin{equation}
 P_Z(j,k)=\tfrac12(p_jq_k+q_jp_k),\qquad
 r_j=\tfrac12(p_j+q_j).
 \label{eqs:pz}
\end{equation}
The four cells with $j,k\in\{0,1\}$ equal $1/(2d)$, the
$4(d-2)$ cells with exactly one index in $\{0,1\}$ equal $1/(4d)$,
and all remaining cells vanish.  The marginal has two entries
$(d+2)/(4d)$ and $d-2$ entries $1/(2d)$.  Substitution into
$\I(P_Z)=\sum_{jk}P_Z(j,k)\log[P_Z(j,k)/(r_jr_k)]$ yields
\begin{equation}
 g_d=\I(Z_A:Z_B)=
 \frac2d\log\frac{8d}{(d+2)^2}
 +\frac{d-2}{d}\log\frac{2d}{d+2}.
 \label{eqs:gd}
\end{equation}
For $d\geq3$, $P_Z$ is nonproduct; for example,
$P_Z(2,2)=0$ but $r_2^2>0$.  Mutual information is a relative entropy,
so $g_d>0$.  This proves the CQC violation for all $d\geq3$ without a
numerical approximation.

For $d=3$, the two Born tables are
\begin{equation}
 P_X=\begin{pmatrix}
 0&1/4&1/4\\ 1/4&0&0\\ 1/4&0&0
 \end{pmatrix},\qquad
 P_Z=\begin{pmatrix}
 1/6&1/6&1/12\\ 1/6&1/6&1/12\\ 1/12&1/12&0
 \end{pmatrix}.
 \label{eqs:qutrit-tables}
\end{equation}
They give
\begin{equation}
 \I_X=1,\qquad
 \I_Z=\frac13\log\frac{3456}{3125}
 =0.048415675908886\ldots .
\end{equation}

\subsection{Full-rank robustness}

Consider
\begin{equation}
 \rho_d^{(\epsilon)}=(1-\epsilon)\rho_d
 +\epsilon\frac{\mathbb I}{d^2},\qquad0<\epsilon<1.
\end{equation}
This state is full rank and separable.  Born probabilities are affine in the
state, and Shannon and von Neumann entropies are continuous in finite
dimension.  Since the gap is strict at $\epsilon=0$, it remains positive on
an open interval.  For the qutrit white-noise path the first numerical zero is
$\epsilon=0.0171610328929\ldots$; this value is descriptive and is not used in
the existence proof.

\section*{Appendix B: The ECQC counterexample}

Let $d$ be an odd prime.  A canonical complete set of $d+1$ MUBs consists of
the computational basis $M_\infty$ and $d$ quadratic bases
$M_a=\{\ket{a,b}:b\in\mathbb F_d\}$, where $\mathbb F_d$ denotes the integers
modulo $d$ and
\begin{equation}
 \ket{a,b}=\frac1{\sqrt d}\sum_{j=0}^{d-1}
 \omega^{aj^2+bj}\ket j,\qquad a,b\in\mathbb F_d.
 \label{eqs:quadratic-mubs}
\end{equation}
Orthonormality within each $a$ follows from the geometric sum over $j$.
Between two distinct $a$ values, the squared magnitude of the quadratic Gauss
sum is $d$, giving overlap $1/d$.  Each $M_a$ is also unbiased to
$M_\infty$ because every computational amplitude has magnitude $1/\sqrt d$.

For ECQC we use a distinct member of the two-branch CC architecture in
Eq.~\eqref{eq:ecqc-branches} of the main text:
\begin{equation}
 \ket{u_d^{\mathrm E}}=\frac{\ket0+\ket1}{\sqrt2},\qquad
 \ket{v_d^{\mathrm E}}=\frac{\ket0-\ket1}{\sqrt2},
\end{equation}
\begin{equation}
 \rho_d^{\mathrm E}=\frac12\proj{u_d^{\mathrm E}}_A
                     \otimes\proj{v_d^{\mathrm E}}_B
                    +\frac12\proj{v_d^{\mathrm E}}_A
                     \otimes\proj{u_d^{\mathrm E}}_B .
 \label{eqs:rho-ecqc}
\end{equation}
The branch states are orthogonal, so $\rho_d^{\mathrm E}$ is again a rank-two
CC state with $\I(A:B)=1$.  In the computational basis their outcome laws
coincide, and hence $\I_{M_\infty}=0$.  In every quadratic MUB their outcome
laws are, up to the common cyclic shift $b\mapsto b+a$,
\begin{equation}
 p_j=\frac{1+c_j}{d},\qquad
 q_j=\frac{1-c_j}{d},\qquad
 c_j=\cos\frac{2\pi j}{d}.
 \label{eqs:ecqc-laws}
\end{equation}
The two amplitudes differ only by the relative sign between their $j=0$ and
$j=1$ components.  The associated same-setting table is
\begin{equation}
 P^{(d)}_{jk}=\frac{1-c_jc_k}{d^2},\qquad
 J_d:=\I(P^{(d)}).
 \label{eqs:ecqc-joint}
\end{equation}
Its marginals are uniform.  Thus the complete-MUB score multiset is
$\{0,J_d,\ldots,J_d\}$, with $d$ copies of $J_d$.  ECQC discards one largest
term and retains the zero term, giving
\begin{equation}
 E_d=(d-1)J_d .
 \label{eqs:Ed}
\end{equation}
At $d=7$, direct evaluation gives
\begin{equation}
 J_7=0.207984126604019\ldots,\qquad
 E_7=1.247904759624115\ldots>1=\I(A:B).
 \label{eqs:d7-ecqc}
\end{equation}
The ancillary verifier constructs all eight bases, checks every pairwise MUB
overlap, reconstructs every Born table, and performs the ECQC minimization.

\subsection{Asymptotic overrun}

Set $\theta_j=2\pi j/d$.  Because the marginals of
Eq.~\eqref{eqs:ecqc-joint} are uniform,
\begin{equation}
 J_d=\frac1{d^2}\sum_{j,k=0}^{d-1}
 \bigl(1-\cos\theta_j\cos\theta_k\bigr)
 \log\!\bigl(1-\cos\theta_j\cos\theta_k\bigr),
 \label{eqs:Jd-sum}
\end{equation}
where $0\log0$ is defined by continuity.  The summand is continuous on the
torus, so
\begin{equation}
 J_d\longrightarrow J_\infty=
 \int_0^{2\pi}\!\int_0^{2\pi}
 \frac{d\theta\,d\phi}{(2\pi)^2}
 \bigl(1-\cos\theta\cos\phi\bigr)
 \log\!\bigl(1-\cos\theta\cos\phi\bigr).
 \label{eqs:Jinf}
\end{equation}
This integral is strictly positive: its joint density is not the product of
its uniform marginals.  Numerically,
\begin{equation}
 J_\infty=0.2067807\ldots .
\end{equation}
Consequently, $E_d=(d-1)J_d$ grows linearly along the infinitely many prime
dimensions, whereas $\I(A:B)=1$ remains fixed.  This proves an unbounded,
rather than isolated, ECQC overrun.

\section*{Appendix C: False quantum-correlation and steering certification}

When Alice measures a basis $\mathcal A=\{\ket{a_i}\}$, outcome $i$ occurs
with probability $p_i$ and prepares a conditional state $\rho_{B|i}$ for Bob.
The Holevo correlation
\begin{equation}
 \chi(\rho|\mathcal A)=S(\rho_B)-\sum_i p_iS(\rho_{B|i})
\end{equation}
upper-bounds the classical information that Bob can extract about Alice's
outcome.  The one-sided quantity of Ref.~\cite{Wu2014} first maximizes this
correlation,
\begin{equation}
 C_1^{\rightarrow}(\rho)=\max_{\mathcal A}\chi(\rho|\mathcal A),
\end{equation}
and then maximizes the Holevo correlation over bases $\mathcal A'$ that are
MUB to any $C_1^{\rightarrow}$-optimal basis:
\begin{equation}
 Q_2^{\rightarrow}(\rho)=
 \max_{\mathcal A\in\operatorname{argmax}C_1^{\rightarrow}}
 \max_{\mathcal A'\perp_{\rm MUB}\mathcal A}
 \chi(\rho|\mathcal A').
 \label{eqs:q2-def}
\end{equation}
The original definition explicitly maximizes over nonunique
$C_1^{\rightarrow}$ bases.

For $\rho_d$, measuring Alice in $X$ identifies the branch perfectly.  Bob's
two branch states are the orthogonal states $\ket u$ and $\ket v$, so the
Holevo quantity is one bit.  Since $S(\rho_B)=1$, this is globally optimal:
$X$ is a $C_1^{\rightarrow}$ basis.  Its MUB $Z$ produces the ensemble whose
Holevo quantity is
\begin{equation}
 h_d=H\!\left(a,a,\underbrace{b,\ldots,b}_{d-2}\right)
 -\frac12\log d-\frac12,\qquad
 a=\frac{d+2}{4d},\quad b=\frac1{2d}.
 \label{eqs:hd}
\end{equation}
Equivalently, $h_d=\I(K:Z_A)$: Bob holds an orthogonal encoding of $K$,
while Alice's $Z$ outcome supplies partial information about it.  At $d=3$,
\begin{equation}
 Q_2^{\rightarrow}(\rho_3)\geq h_3
 =0.190874504621109\ldots>0.
\end{equation}
Thus the explicit conclusion of Ref.~\cite{Wu2014} that complementary-basis
quantum correlation vanishes on every CQ, and hence every CC, state is false.

The same state directly tests the universal 1SSDI claim of
Ref.~\cite{Jebarathinam2025}.  In the formulation at issue, the data would
certify steering only if they admitted no local-hidden-variable--local-hidden-
state (LHV--LHS) explanation whose classical hidden dimension obeys the stated
bound.  For an arbitrary measurement $x$ on Alice, the resulting collection
of subnormalized conditional states for Bob (the assemblage) is
\begin{equation}
 \sigma_{a|x}=\frac12 p(a|x,u)\proj v
              +\frac12 p(a|x,v)\proj u.
 \label{eqs:lhs}
\end{equation}
Equation~\eqref{eqs:lhs} is an LHV--LHS model with $\lambda\in\{0,1\}$,
probabilities $1/2$, response
functions $p(a|x,u)$ and $p(a|x,v)$, and hidden states $\proj v$ and
$\proj u$.  Its hidden-variable dimension is $d_\lambda=2\leq d_A$ for
all $d\geq3$, satisfying the stated 1SSDI restriction.  Hence the state cannot
demonstrate 1SSDI steering under that definition despite
$Q_2^{\rightarrow}>0$.  Its operator-Schmidt rank is also exactly two, because
Eq.~\eqref{eqs:rho} contains two linearly independent projectors on each side.
It therefore simultaneously contradicts the claimed equivalence
$Q_2^{\rightarrow}=0\Leftrightarrow L_R\leq d_{\min}$.

\section*{Appendix D: Conditional-probability counterexamples}

For a basis $M$, let $a_i$ and $b_i$ be Alice's and Bob's outcomes assigned
the same label $i$.  Reference~\cite{Maccone2015} defined, with this fixed
same-index pairing,
\begin{equation}
 \mathcal S_M=\sum_{i=0}^{d-1}p(a_i|b_i)
\end{equation}
and conjectured the separable-state interval
\begin{equation}
 1\leq\mathcal S_X+\mathcal S_Z\leq d+1.
 \label{eqs:S-interval}
\end{equation}
The qutrit tables in Eq.~\eqref{eqs:qutrit-tables} have strictly positive
Bob marginals.  Their diagonals give
\begin{equation}
 \mathcal S_X=0,\qquad \mathcal S_Z=\frac45,
\end{equation}
which violates the lower bound of Eq.~\eqref{eqs:S-interval} with the same
qutrit CC state used for CQC and $Q_2^{\rightarrow}$.

The original article noted that outcome pairing is arbitrary and suggested
maximizing or minimizing over permutations, but evaluated the fixed pairing
for simplicity.  Consider the natural setting-wise maximum
\begin{equation}
 \mathcal S_M^\star=\max_{\pi\in S_d}
 \sum_{j=0}^{d-1}p(a_{\pi(j)}|b_j).
 \label{eqs:S-star}
\end{equation}
The upper bound $d+1$ still fails.  We give an explicit $d=4$ separable
construction.  Let $\{\ket{z_j}\}$ be the computational basis and
$\ket{x_k}=2^{-1}\sum_{j=0}^3 i^{jk}\ket{z_j}$ its Fourier MUB.  Define
four product components $\ket{a_\ell}\ket{b_\ell}$ by
\begin{align}
 \ket{a_0}&=\ket{x_1},&
 \ket{b_0}&=\frac{(\mathbb I-\proj{z_1})\ket{x_2}}
 {\|(\mathbb I-\proj{z_1})\ket{x_2}\|},\\
 \ket{a_1}&=\frac{\ket{z_2}-\ket{x_3}}{\sqrt3},&
 \ket{b_1}&=\frac{\ket{z_0}+\ket{z_2}+2\ket{z_3}}{\sqrt6},\\
 \ket{a_2}&=\frac{\ket{z_1}+\ket{z_3}}{\sqrt2},&
 \ket{b_2}&=\frac{\ket{z_0}-\ket{z_2}}{\sqrt2},\\
 \ket{a_3}&=\ket{z_0},&
 \ket{b_3}&=\frac{(\mathbb I-\proj{x_2})\ket{z_1}}
 {\|(\mathbb I-\proj{x_2})\ket{z_1}\|},
\end{align}
and
\begin{equation}
 \tau_4=\sum_{\ell=0}^3w_\ell
 \proj{a_\ell}\otimes\proj{b_\ell},
 \qquad (w_0,w_1,w_2,w_3)=\frac1{100}(3,15,79,3).
 \label{eqs:tau4}
\end{equation}
Exhausting all $4!$ assignments in both settings gives
\begin{equation}
 \mathcal S_Z^\star+\mathcal S_X^\star
 =\frac{40846}{7785}=5.246756583172768\ldots>5.
\end{equation}
The maximizing zero-based permutations are $(3,0,1,2)$ for $Z$ and
$(3,2,1,0)$ for $X$.  Mixing $1\%$ white noise,
\begin{equation}
 \tau_4'=0.99\tau_4+0.01\frac{\mathbb I}{16},
\end{equation}
produces a full-rank separable state and still gives
\begin{equation}
 \mathcal S_Z^\star+\mathcal S_X^\star
 =5.071079396064217\ldots>5.
\end{equation}
Thus neither a fixed-label lower bound nor the natural
setting-wise-permutation upper bound is a universal separability criterion.

\section*{Appendix E: Two-branch classical null test and scope}

The qutrit experiment requires no entangling gate.  A source draws a fair bit
$K$ and prepares $\ket u_A\ket v_B$ for $K=0$ or
$\ket v_A\ket u_B$ for $K=1$.  Both parties independently and randomly choose
the $X$ or $Z$ analyzer after preparation.  The ideal same-setting tables are
Eq.~\eqref{eqs:qutrit-tables}, with targets
\begin{equation}
 \I_X=1,\quad \I_Z=0.0484156759\ldots,\quad
 \mathcal S_X=0,\quad \mathcal S_Z=0.8.
\end{equation}
Conditional tomography of Bob's two-dimensional support additionally gives
$Q_2^{\rightarrow}\geq0.1908745046\ldots$.  This source can therefore be run
through the exact settings and processing of a proposed multibasis
certification protocol.  Firing on it is a direct classical false positive.

The diagnostic has four fixed steps: (i) program the source in
Eq.~\eqref{eqs:rho}; (ii) draw $K$ and emit the product pair before the analyzer
settings are chosen; (iii) run the candidate protocol with its original
measurements, label conventions, decoders, loss treatment, abort rules, and
postselection; and (iv) evaluate its advertised score and resource threshold.
The same two-branch source is used for every setting.  No entangling gate,
tomographic reconstruction, or optimization over all states declared
classical or separable is required.  If the protocol crosses its threshold, it
has a direct classical false positive.

Passing the test is only a necessary diagnostic.  A complete resource witness
must still optimize jointly over the full class of states that it declares
classical or separable and all processing allowed by its operational model; a
sum of separately justified setting-wise bounds is not a substitute.  The
contribution here is the explicit minimal classical challenge source and the
common failure mode it exposes, not a new general optimization algorithm.

Finally, the counterexamples do not invalidate entropic-uncertainty relations
or the independently proved mutual-information entanglement criterion of
Ref.~\cite{Maccone2015}.  Other MUB witnesses whose separable bounds have
independent proofs also remain intact \cite{Spengler2012,Coles2017}.  The
counterexamples invalidate only the exact universal inequalities and
operational implications evaluated above.  In particular, they do not imply a
generic attack on QKD, device-independent security, or finite-key analysis.
Any cryptographic consequence must be rederived inside its specified source,
side-information, and adversary model.

\bibliography{cqc_counterexample}

\end{document}